\documentclass[fleqn,multphys,vecphys]{svmult}

\usepackage{makeidx}   
\usepackage{graphicx}  
\makeindex             

\newcommand{\iDev}[1]{#1}
\newcommand{\iName}[1]{#1}
\newcommand{\iStreet}[1]{#1}
\newcommand{\iPostcode}[1]{#1}
\newcommand{\iCity}[1]{#1}
\newcommand{\iCountry}[1]{#1}
\newcommand{\email}[1]{\texttt{#1}}

\def\abstractname{Abstract.}%

\begin{document}

\title*{Spin dynamics in high-mobility two-dimensional electron systems}          
\toctitle{Spin dynamics in high-mobility two-dimensional electron systems}        
\titlerunning{Spin dynamics in high-mobility 2D electron systems}                                

\author{Tobias Korn \inst{1}                     
\and Dominik Stich \inst{1} \and Robert Schulz \inst{1} \and Dieter
Schuh \inst{1}  \and Werner Wegscheider \inst{1} \and Christian
Sch\"uller \inst{1} }
\authorrunning{T. Korn et al.} 

\institute{\iDev{Institut f\"ur Experimentelle und Angewandte Physik},                  
\iName{Universit\"at Regensburg}, \iStreet{Universit\"atsstrasse
31}, \iPostcode{93040}, \iCity{Regensburg},
\iCountry{Germany},\newline
\email{tobias.korn@physik.uni-regensburg.de}}

\maketitle

\begin{abstract}
Understanding the spin dynamics in semiconductor heterostructures is
highly important for future semiconductor spintronic devices. In
high-mobility two-dimensional electron systems (2DES), the spin
lifetime strongly depends on the initial degree of spin polarization
due to the electron-electron interaction. The Hartree-Fock (HF) term
of the Coulomb interaction acts like an effective  out-of-plane
magnetic field and thus reduces the spin-flip rate. By time-resolved
Faraday rotation (TRFR) techniques, we demonstrate that the spin
lifetime is increased by an order of magnitude as the initial spin
polarization degree is raised from the low-polarization limit to
several percent. We perform control experiments to decouple the
excitation density in the sample from the spin polarization degree
and investigate the interplay of the internal HF field and an
external perpendicular magnetic field. The lifetime of spins
oriented in the plane of a [001]-grown 2DES is strongly anisotropic
if the Rashba and Dresselhaus spin-orbit fields are of the same
order of magnitude. This anisotropy, which stems from the
interference of the Rashba and the Dresselhaus spin-orbit fields, is
highly density-dependent: as the electron density is increased, the
kubic Dresselhaus term becomes dominant und reduces the anisotropy.
\end{abstract}
\section{Introduction}
In recent years, semiconductor spintronics
\cite{Awschalom1,Fabian04,Fabian07} research has found increased
interest, in part due to new materials like ferromagnetic
semiconductors \cite{Ohno1996}. Among the key requirements for
semiconductor spintronic devices is an understanding of the spin
dephasing mechanisms in semiconductors. For GaAs, many experimental
studies have focused on slightly n-doped bulk material, where
extremely long spin lifetimes ($\geq 100$\,ns) were observed
\cite{Kikkawa1, Kikkawa2} for doping levels close to the
metal-insulator transition. Even though this material has low
mobility and poor conductivity properties, it was used in a number
of spin injection \cite{Crooker} and transport \cite{Kikkawa3, Aw07}
experiments.

Relatively few  studies have been performed on high-mobility
two-dimensional electron systems (2DES): Brand et al. investigated
the weak scattering regime and the temperature dependence of the
D'Yakonov-Perel (DP) mechanism  in a high-mobility 2DES
\cite{Brand}, while Leyland et al. \cite{Leyland07} experimentally
showed the importance of electron-electron collisions for spin
dephasing. Several groups performed experiments on [110]-grown
quantum wells and 2DES, in which the DP mechanism can be suppressed
for spins aligned along the growth direction \cite{Ohno99}, while it
remains active for other spin orientations \cite{wugo,Dohrmann}. In
these systems, gate control of the spin lifetime was demonstrated by
Karimov et al. \cite{Karimov}. Theoretical studies investigating the
electron-electron interaction in 2DES were performed using
perturbation theory by Glazov and Ivchenko \cite{Ivchenko}, while Wu
et al. developed a powerful microscopic many-body approach
\cite{wu_JAP03} to study 2DES far from thermal equilibrium,
considering all relevant scattering mechanisms, including
electron-hole and electron phonon scattering \cite{wu1,wu2,wu3,wu4}.
\section{Theory}
\subsection{Optical orientation of electrons in a 2D
electron system} In  GaAs, the conduction band has \emph{s}-like
character, while the valence bands have \emph{p}-like character. The
light-hole (LH $ J_z=\pm \frac{1}{2}$) and the heavy-hole (HH $
J_z=\pm \frac{3}{2}$) valence bands are degenerate at \emph{k}=0 for
bulk GaAs. Optical excitation   above the bandgap can generate
electron-hole pairs through photon absorption. Due to angular
momentum conservation, a circularly-polarized photon (s=1) may
excite both, electrons with spin up (from a  light-hole valence band
state with $J_z=- \frac{1}{2}$) and spin down (from  a heavy hole
valence band state with $J_z=- \frac{3}{2}$). Due to the different
 probabilities of these transitions, a finite spin polarization may
 be created in bulk GaAs \cite{OptOr}. In a quantum well, the
 $\vec{k}=0$
 degeneracy of the light and heavy hole valence bands is lifted due
 to confinement. By resonantly exciting only the HH or LH transition
 in a quantum well (QW) with circularly-polarized light, almost
 100\,percent spin polarization may be generated \cite{MarieCentPourCent}.
 If a 2DES is created within the QW by
 (modulation) doping, the electrons occupy conduction band states up
 to the Fermi energy. Optical absorption is  only possible into
 states above the Fermi energy, and  does not occur at $\vec{k}=0$.
 For $\vec{k}\neq$0, the valence band states in a QW consist
 of an admixture of heavy-hole and light-hole states and
 $J_z$ is no longer a good quantum number \cite{Pfalz}. Therefore,
 excitation with circularly polarized light will yield a mixture of
 spin-up and spin-down electrons in the conduction band, depending
 on the excitation wavelength. Additionally, the 2D electron system
 is typically unpolarized in thermal equilibrium.
 In order to create a large  spin polarization in a 2D
 electron system by optical excitation, the excitation density
 therefore has to be on the order of the 2DES density if we assume that the 2D system
 returns to thermal equilibrium in between excitation pulses. The initial
 spin polarization degree \emph{P}, created by a short optical pulse, may be calculated using the following
 formula:
 \begin{equation}\label{IniPol}
    P=\frac{n_{ph}}{n_e+n_{ph}^{tot}}\;.
 \end{equation}
Here, $n_{ph}=\xi \cdot n_{ph}^{tot}$ is the spin-polarized fraction
$\xi$ of the optically created electron density, $n_{ph}^{tot}$ is
the total electron density that is optically created, and $n_e$ is
the background electron density of the 2DES.
\subsection{Rashba and Dresselhaus spin-orbit fields in a [001]
quantum well} 
\begin{figure}
\includegraphics[width=11cm]{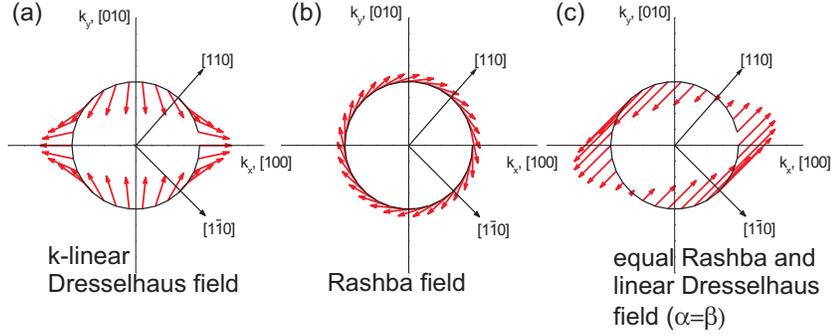}
    \caption{ (a) Symmetry of the linear Dresselhaus field in a 2D quantum well. The red
    arrows
    indicate the effective field direction and amplitude depending on the electron
    \emph{k} vector in the \emph{xy} plane. (b)Symmetry of the Rashba field in a 2D
    quantum well. (c) Vector
    sum of linear Dresselhaus and Rashba field (combined effective field, CEF) for
    identical amplitudes
    ($\ualpha=\ubeta$).}
      \label{Rashba_Dressel_2Panel}
\end{figure}

In  crystal structures which lack  inversion symmetry,
like GaAs, the spin-orbit interaction may be described by an
intrinsic, $\vec{k}$-dependent magnetic field $\vec{B}_i(\vec{k})$,
which causes a precession of the electron spin. Typically, a Larmor
precession frequency vector corresponding to the electron precession
about this internal field is defined as
$\vec{\Omega(k)}=\frac{g}{\hbar}\cdot \umu_B \vec{B}_i(k)$. In bulk
structures, $\vec{\Omega(k)}$ is
 cubic in $\vec{k}$ and has the following form \cite{bia}:
\begin{equation}\label{Bulk_BIA}
\vec{\Omega(k)}_{BIA}=\frac{\gamma}{\hbar}\cdot
[k_x(k_y^2-k_z^2),k_y(k_z^2-k_x^2),k_z(k_x^2-k_y^2)]\;.
\end{equation} This term stems from the inversion asymmetry of the
crystal lattice and is therefore often called bulk inversion
asymmetry (BIA) term or Dresselhaus term. In a quantum well grown
along the \emph{z} direction, the momentum along the growth
direction is quantized due to confinement. In first approximation,
the expectation value is $\langle k_z^2\rangle = (\pi / d)^2$, where
\emph{d} is the quantum well thickness. It follows that :
\begin{equation}\label{2D_BIA}
\vec{\Omega(k)}_{BIA(2D)}=\frac{\gamma}{\hbar}\cdot
[k_x(k_y^2-\langle k_z^2\rangle),k_y(\langle
k_z^2\rangle-k_x^2),0]\;.
\end{equation}
Typically in a 2D electron system, the in-plane momentum
$(k_\parallel)^2$ is  smaller than $\langle k_z^2\rangle$.
Therefore, terms kubic in the in-plane momentum are often neglected,
resulting in the following approximation:
\begin{equation}\label{lin_Dressel}
\vec{\Omega(k)}_{Dressel}=\frac{\ubeta}{\hbar}\cdot [-k_x,k_y,0]\;.
\end{equation}
This is linear in the in-plane momentum $k_\parallel$ and therefore
typically called linear Dresselhaus term. Its symmetry is shown in
Fig. \ref{Rashba_Dressel_2Panel} (a). Additionally, a lack of
structure inversion symmetry along the quantum well growth direction
causes a second intrinsic effective magnetic field, the so-called
Rashba field \cite{sia}.
\begin{equation}\label{Rashba}
\vec{\Omega(k)}_{Rashba}=\frac{\ualpha}{\hbar}\cdot [k_y,-k_x,0]\;.
\end{equation}
Structure inversion asymmetry (SIA) along the growth direction may
be induced by different barrier materials on either side of the
quantum well, single-sided or asymmetric modulation doping resulting
in an effective electric field due to ionized donors, or the
application of an external electric field by a gate voltage. The
symmetry of the Rashba field is shown in Fig.
\ref{Rashba_Dressel_2Panel} (b).
\subsection{Spin dephasing/relaxation}
The main mechanism for spin dephasing in GaAs bulk and quantum well
structures at low temperatures is the so-called D'Yakonov-Perel
mechanism \cite{dp}. It is caused by the \emph{k}-dependent
spin-orbit fields which cause a precession of the electron spin. In
an ensemble of electrons, the \emph{k} values are distributed
according to Fermi statistics, causing different precession
frequencies and directions for the electron spins, which leads to
dephasing. Two regimes can be distinguished according to the
relationship between the average precession frequency $\bar{\Omega}$
and the momentum relaxation time $\tau_P$:
\begin{enumerate}
  \item weak scattering regime ($\bar{\Omega} \cdot \tau_P >1$):
  here, the electron spins may precess more than one full cycle
  about the spin-orbit field before they are scattered and their \emph{k}
  value changes. A collective precession of electron spins may be
  observed in this regime.
  \item strong scattering regime ($\bar{\Omega} \cdot \tau_P < 1$):
  here, the electrons are scattered so frequently that the
  spin-orbit field acts like a rapid fluctuation. In this regime,
  the spin relaxation time $T_2^*$ is inversely proportional to the momentum
  relaxation time: $\frac{1}{T_2^*} \propto \tau_P$.
\end{enumerate}
\subsection{Magneto-Anisotropy}
In samples where both the Rashba and the Dresselhaus terms are
present and of the same magnitude, the effective spin-orbit field
has to be calculated as the vector sum of the two terms:
\begin{equation}\label{CEF}
\vec{\Omega(k)}_{CEF}=\frac{1}{\hbar}\cdot [(\ualpha k_y- \ubeta
k_x),(\ubeta k_y- \ualpha k_x),0]\;.
\end{equation}
Due to the different symmetry of the two contributions, the
resulting combined effective field (CEF) may show a well-defined
preferential direction if $\ualpha= \ubeta$:
\begin{equation}\label{CEF_alpha}
\vec{\Omega(k)}_{CEF}=\frac{\ubeta}{\hbar}\cdot [(k_y-k_x),(k_y-
k_x),0]\equiv \frac{\ubeta}{\hbar} (k_y-k_x)\cdot [1,1,0]\;.
\end{equation}
In this case, the CEF points along the in-plane [110] direction for
any value of $\vec{k}$, as Fig. \ref{Rashba_Dressel_2Panel} c)
shows. (For $\ualpha= -\ubeta$ this preferential direction becomes
[1$\bar{1}$0]). This means that electron spins that point along the
[110] direction experience no torque and can therefore not dephase
due to the D'Yakonov-Perel mechanism, which is effectively blocked
for this spin orientation, while electron spins pointing along
[1$\bar{1}$0] or along the growth direction will experience a torque
and start dephasing. This spin dephasing anisotropy was first
pointed out by Averkiev and Golub \cite{Averkiev}, and recently
observed experimentally \cite{Averkiev06, Liu}.
\section{sample structure and preparation}
Our sample was grown by molecular beam epitaxy on a $[001]$-oriented
semi-insulating GaAs substrate. The active region is a 20 nm-wide,
one-sided modulation-doped GaAs-Al$_{0.3}$Ga$_{0.7}$As single QW.
The electron density and mobility at $T=4.2$\,K are $n_e=2.1\times
10^{11}$\,cm$^{-2}$ and $\umu_e=1.6\times 10^6$\,cm$^2$/Vs,
respectively. These values were determined by transport measurements
on an unthinned sample. For measurements in transmission geometry,
the sample was glued onto a sapphire substrate with an optical
adhesive, and the substrate and buffer layers were removed by
selective etching.
\section{Measurement techniques}
\subsection{Time-resolved Kerr/Faraday rotation}
For both, the time-resolved Faraday rotation (TRFR) and the
time-resolved Kerr rotation (TRKR) measurements, two laser beams
from a mode-locked Ti:Sapphire laser, which is operated at 80\,MHz
repetition rate, were used. The laser pulses had a temporal length
of about 600\,fs each, resulting in a spectral width of about
3-4\,meV, which allowed for near-resonant excitation. The laser
wavelength was tuned to excite electrons from the valence band to
states slightly above the Fermi energy of the host electrons in the
conduction band. Both laser beams were focused to a spot of
approximately 60\,$\umu$m diameter on the sample surface. The pump
pulses were circularly polarized by an achromatic
$\frac{\lambda}{4}$ plate in order to create spin-oriented electrons
in the conduction band, with spins aligned perpendicular to the QW
plane. The weaker probe pulses were linearly polarized. The
polarization rotation of the transmitted/reflected probe beam was
analyzed by an optical bridge detector. In order to separate the
time evolution of the spin polarization from the photocarrier
dynamics, all measurements were performed using both helicities for
the circularly-polarized pump beam \cite{stich_PhysE08}. The TRFR
measurements were performed in a split-coil magnet cryostat with a
$^3$He insert, allowing for sample temperatures between 1.5\,K and
4.5\,K. The TRKR measurements were performed in a continuous-flow He
cold finger cryostat. In this cryostat, non-thinned samples from the
same wafer were used. Unless otherwise stated, the experiments were
carried out at a nominal sample temperature of T=4.5\,K.
\section{Experimental results}
\subsection{Variation of spin polarization degree}
\begin{figure}
\includegraphics[width=8.5cm]{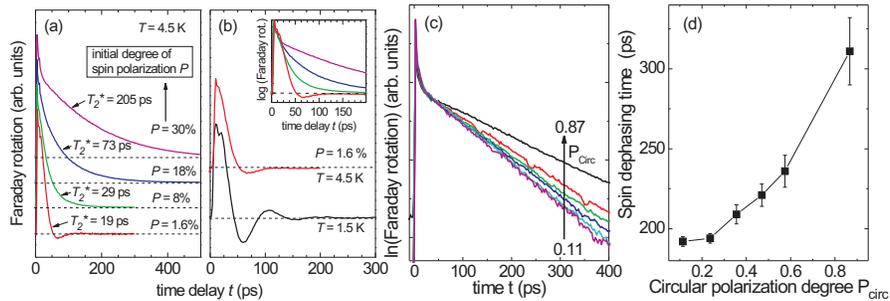}
    \caption{(a) Normalized TRFR traces for different pump beam fluence and therefore
    different    initial spin polarization.
    (b) Comparison of TRFR traces for low initial spin polarization at different
    temperatures.
    The inset shows the data from (a) with a log scale.
    (c) TRKR traces for varying circular polarization degree of the pump laser
    beam. The curves have been normalized for easier comparison. Note the log scale.
    (d) Spin    relaxation time as a function of circular polarization degree of the
    pump beam, extracted    from the TRKR traces.
    ((a,b) reprinted  with permission
    from \cite{stich_PRL07}.
     Copyright (2007) by the American Physical Society.)}
      \label{Zero_2Panel}
\end{figure}

Typically, in time-resolved Faraday/Kerr rotation measurements on
semiconductor heterostructures, the excitation density is kept very
low to avoid heating the sample. Here, we present experiments in
which the excitation density was considerable, resulting in a
significant initial spin polarization. Figure \ref{Zero_2Panel} (a)
shows a series of TRFR measurements performed without external
magnetic field. In this measurement, the excitation density, and
therefore the initial spin polarization, was increased by increasing
the laser pump fluence. The curves typically show a biexponential
decay of the spin polarization, except for the lowest curve, where a
strongly damped oscillation is visible. The oscillatory behavior
will be discussed in the following subsection.  We associate the
fast decay with the spin dephasing of the photoexcited holes, which
typically lose their spin orientation within a few picoseconds, and
the slower decay with the spin dephasing time of the electrons,
$T_2^*$. It is clearly seen how  $T_2^*$ is increased by more than
an order of magnitude as the initial spin polarization is increased
from the low-polarization limit to about 30\,percent. This increase
is due to the Hartree-Fock (HF) term of the Coulomb interaction (as
predicted by Weng and Wu \cite{wu1}), which acts as an out-of-plane,
$\vec{k}$-dependent effective magnetic field. This effective field
lifts the degeneracy of the spin-up and spin-down states, causing a
spin-flip to require a change in energy. This significantly reduces
the spin-flip rate and increases the spin dephasing time. In
microscopic calculations including the HF term of the Coulomb
interaction, the measurements shown in Fig. \ref{Zero_2Panel} (a)
could be reproduced with great accuracy.

In order to clearly demonstrate that the observed effect is due to
the increase in spin polarization, and not caused by either a change
in electron density or sample heating, we performed control
experiments in which the initial spin polarization was varied while
the excitation density was kept constant \cite{Stich_PRB07}. In
order to achieve this, the circular polarization degree of the pump
laser pulse was varied from almost 100\,percent circular
polarization to almost linear polarization, by gradual rotation of
the $\frac{\lambda}{4}$ plate in the pump beam. By this means, the
fraction of spin-polarized electrons created by the pump beam could
be tuned. Figure \ref{Zero_2Panel} (c) shows a series of TRKR
measurements performed in this way. It is clearly visible how an
increase of the circular polarization degree of the pump beam, and
therefore an increase of the initial spin polarization, leads to
reduced spin dephasing and therefore a shallower slope in the
logarithmic plot of the TRKR trace. The TRKR traces have been
normalized for easier comparison. In Fig. \ref{Zero_2Panel} (d), the
spin dephasing times extracted from the measurement series are
plotted as a function of the circular polarization degree.
\subsection{Coherent zero-field oscillation}
In Fig. \ref{Zero_2Panel} (b), two TRFR traces taken without
external magnetic field at low excitation density are shown. The
upper (red) curve, showing a strongly damped oscillation, represents
the same data as the lowest (red) curve in Fig. \ref{Zero_2Panel}
(a). The lower curve in \ref{Zero_2Panel} (b) was taken using the
same excitation density, but at a lower sample temperature of
1.5\,K. Here, the damping is significantly reduced. Both traces are
representative of the weak scattering regime of the DP mechanism: an
ensemble of electrons with $\vec{k}$ slightly above the Fermi wave
vector is generated by the pump pulse. While the orientation of the
effective spin orbit field for these electrons depends on the
electron $\vec{k}$ vector, it generally lies in the sample plane,
leading to spin precession. The \emph{z} component of the electron
spins  oscillates, and the damped oscillation we observe, is the
coherent superposition of these oscillations. We note that this
effect was first reported by Brand et al. \cite{Brand}. Two effects
contribute to the damping of this coherent oscillation:

1. The amplitude of the effective spin-orbit field depends on
$\vec{k}$ if both the Rashba and the Dresselhaus terms are present
(see for example Fig. \ref{Rashba_Dressel_2Panel} c)), leading to a
(\emph{reversible}) dephasing of the electron spins due to different
precession frequencies.

2. Initially, the electron spins are aligned along the growth
direction by the laser pump pulse, therefore the effective
spin-orbit field is perpendicular to the electron spin for the whole
ensemble. Precession tilts the electron spins into the sample plane,
and momentum scattering changes $\vec{k}$ and $\vec{\Omega(k)}$,
leading to different angles between the electron spins and the
effective spin-orbit fields. This causes (\emph{irreversible})
dephasing due to different spin precession frequencies. Momentum
scattering is temperature-dependent, hence the damping of the
coherent oscillation becomes more pronounced as the sample
temperature is raised.
\subsection{Spin dephasing in an external magnetic field}
\begin{figure}
\includegraphics[width=10cm]{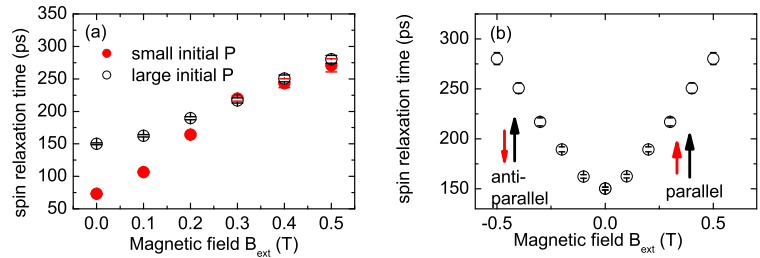}
    \caption{(a) Spin dephasing times as a function of an external magnetic field perpendicular
    to the QW plane for small and large initial spin polarization. (b) Same as (a) for large
    initial spin polarization and both polarities of the external magnetic field.}
      \label{Feld_2Panel}
\end{figure}
Here, we present experiments performed in small external magnetic
fields applied perpendicular to the quantum well plane. The field
amplitude was kept below 0.5\,Tesla, where effects due to Landau
quantization may be neglected. In this regime, we observe a
competition between the Hartree-Fock effective out-of-plane field,
created by the interaction between spin-aligned electrons, and the
external magnetic field. As Fig. \ref{Feld_2Panel} (a) shows, the
application of an external magnetic field monotonically increases
the spin dephasing time for both small and large initial spin
polarization. At zero and small external fields, the HF field
dominates, leading to a significantly longer spin lifetime for large
initial spin polarization. However, as the external field amplitude
is increased, the spin lifetimes for large and small initial spin
polarization start to merge and become almost identical for magnetic
fields of 0.4\,Tesla and above. This indicates that the mean value
of the HF field for the large initial spin polarization is below
0.4\,Tesla.

 The spin polarization, which we create optically, defines a
preferential direction along the growth axis. The external magnetic
field can therefore be applied parallel or antiparallel to this
preferential direction. Figure \ref{Feld_2Panel} (b) shows that for
both external field directions the spin lifetime increases
monotonically, and the curve is symmetric in the external field B.
This indicates that the external magnetic field cannot be used to
compensate the effective Hartree-Fock field, since that is
$\vec{k}$-dependent.
\subsection{Probing magneto-anisotropy}
\begin{figure}
\includegraphics[width=10cm]{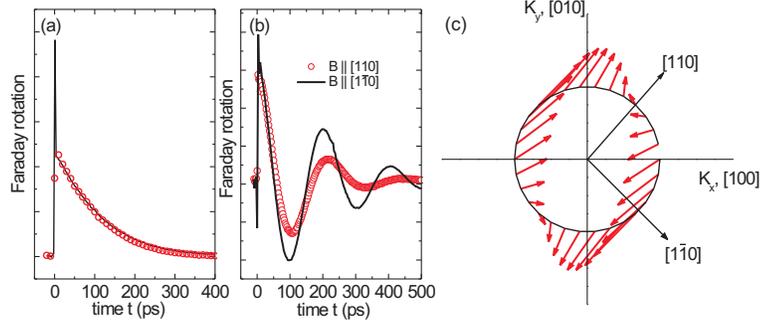}
    \caption{(a) Time-resolved Faraday rotation traces for two orientations of the
     sample without external
    magnetic field. (b)TRFR traces at 1\,Tesla in-plane magnetic field, for two
    relative orientations of the
    sample and the  magnetic field. (c) Symmetry of the combined effective field for
    Rashba-Dresselhaus ratio
    $\ualpha = 0.65 \ubeta$.}
      \label{MagAn_2Panel}
\end{figure}

The magneto-anisotropy in a [001]-grown GaAs QW leads to different
spin lifetimes for electron spins aligned along different in-plane
directions. In our experimental setup, however, we create an
out-of-plane spin polarization by the circularly-polarized pump
pulse, and the detection scheme using the linearly-polarized probe
beam at near-normal incidence is only sensitive to the out-of-plane
component of the spin polarization. Therefore, in order to probe the
in-plane magneto-anisotropy, we use an in-plane magnetic field to
force the spins to precess into the sample plane and back out again.
By this means, the spin lifetimes we observe, represent an average
of the out-of-plane and the different in-plane spin lifetimes within
the sample and therefore allow us to infer the in-plane anisotropy.
In the experiment, we use two sample pieces from the same wafer,
which are mounted in the cryostat with their in-plane [110] axis
either parallel or perpendicular to the applied in-plane magnetic
field. Figure \ref{MagAn_2Panel} shows the experimental results: (a)
without an applied magnetic field, both sample pieces (black line
and red circles) show the same spin lifetime $T_2^*= 113 \pm 1$\,ps,
as we expect due to symmetry reasons. (b) As an in-plane magnetic
field of 1\,Tesla is applied, the spin lifetimes become markedly
different: if the magnetic field is applied along the in-plane [110]
direction, the electron spins are forced to precess into the
[1$\bar{1}$0] direction. The spin lifetime average between the [001]
and the [1$\bar{1}$0] is $T^*_{2[1\bar{1}0]}= 127 \pm 1$\,ps. If,
however, the magnetic field is applied along [1$\bar{1}$0], forcing
the electron spins into the [110] direction, the averaged spin
lifetime $T^*_{2[110]}= 204 \pm 2$\,ps increases by about
60\,percent. This is a clear indication of the in-plane
magneto-anisotropy. From the magnetic-field dependence of the
averaged spin lifetimes, both the Rashba and the Dresselhaus
coefficients can be determined by comparing the experimental data to
numerical many-body calculations \cite{Stich_PRBRC07,korn_PhysE08}.
Here, we find values of $\ubeta= 1.38$\,meV${\AA}$ and $\ualpha=
0.9$\,meV${\AA}$, yielding a ratio $\frac{\ualpha}{\ubeta}=0.65$.
The symmetry of the CEF for this ratio is shown in Fig.
\ref{MagAn_2Panel} (c): the preferential direction of the CEF along
the [110] direction is clearly visible. Using these values, an
in-plane spin lifetime anisotropy of 60 to 1 can be inferred from
the calculations. Thus, the spin lifetime for electron spins
initially \emph{aligned} along the [110] axis is estimated to be
about 8\,ns. We note that in order to observe the magneto-anisotropy
by spin precession, the initial spin polarization has to be so high
that the spin lifetime is long enough to allow for at least one
precession cycle.
\subsection{Breakdown of magneto-anisotropy}
\begin{figure}
\includegraphics[width=11cm]{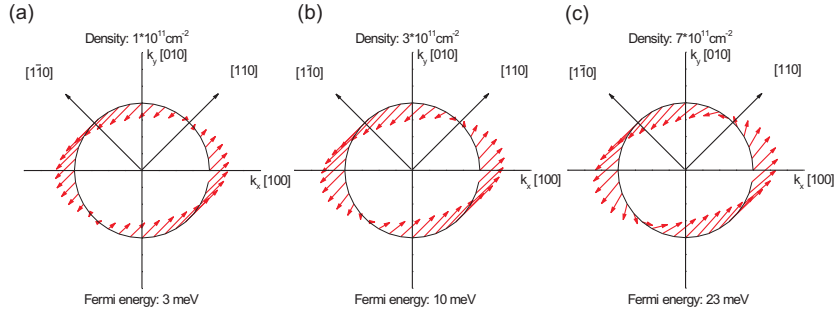}
    \caption{Calculated symmetry of the combined effective field due to Rashba
    and \emph{kubic}
    Dresselhaus terms. The carrier density and thus the Fermi energy is increased
    from (a) to (c).}
      \label{MagAn_breakdown}
\end{figure}
The spin-dephasing anisotropy in [001]-grown QWs is an approximation
based on the interference of the Rashba and the \emph{linear}
Dresselhaus spin-orbit fields. In this approximation, the vector sum
of Rashba and Dresselhaus fields points along the [110] direction
for $\ualpha = \ubeta$, regardless of $\vec{k}$. As the carrier
density in a 2DES is increased, the kubic Dresselhaus term has to be
taken into account, and the symmetry of the combined effective field
(CEF) changes. Here, we show density-dependent calculations of the
symmetry of the CEF as a function of the carrier density. The
calculations were performed for a 20\,nm wide quantum well, and for
the calculations the ratio of the Rashba and Dresselhaus terms was
kept constant at $\ualpha = \ubeta$. As Fig. \ref{MagAn_breakdown}
(a) shows, for low density the orientation of the CEF remains almost
perfectly aligned along the [110] direction. As soon as the density
is increased to values more common for a 2DES, as in
\ref{MagAn_breakdown} (b) and (c), a deviation from this orientation
can be observed for a range of $\vec{k}$ values. In order to
quantify the breakdown of the magneto-anisotropy, we calculate the
average torque $\bar{\vec{\tau}}$ acting on an electron spin
\emph{s} pointing along the in-plane $[110]$ or $[1\bar{1}0]$
directions: $|\bar{\vec{\tau}}|=\sum_k|\vec{s}\times
\vec{B}_{eff}(\vec{k})|.$ Here, the sum is over all directions of
$\vec{k}$ for $|\vec{k}|=k_F$. From these values, we can determine
the (\emph{dimensionless}) anisotropy
$\vec{\bar{\tau}_{[1\bar{1}0]}}/ \vec{\bar{\tau}_{[110]}}$of the
initial torque acting on electron spins pointing along $[110]$ and
$[1\bar{1}0]$,  and track its density-dependence.
\begin{figure}
\includegraphics[width=5cm]{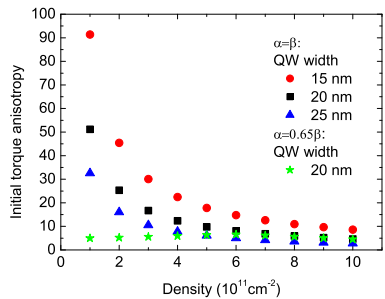}
    \caption{Anisotropy of the initial torque acting on electron spins
    pointing along the
    in-plane $[110]$ and $[1\bar{1}0]$ directions as a function of electron density.
     Calculations for
    15\,nm (red circles), 20\,nm (black squares), and 25\,nm (blue triangles) wide
     QWs with equal Rashba and
    Dresselhaus terms ($\ualpha = \ubeta$) are shown, as well as the values for
    a 20\,nm wide QW
    with ($\ualpha =0.65 \ubeta$) (green stars).}
      \label{torque_ani}
\end{figure}
Figure \ref{torque_ani} shows the results of this calculation for
15-25\,nm wide quantum wells with equal Rashba and Dresselhaus terms
($\ualpha = \ubeta$). It can be clearly seen how the torque
anisotropy is reduced by an order of magnitude as the density is
increased from values common for low-doped 2DES samples ($n=1\times
10^{11}$\,cm$^{-2}$) to highly-doped 2DES ($n=8-10\times
10^{11}$\,cm$^{-2}$). For comparison, the same calculation was
performed for a 20\,nm wide QW with ($\ualpha =0.65 \ubeta$), the
ratio corresponding to the sample investigated in our experiments.
There, the torque anisotropy is significantly lower than for
($\ualpha = \ubeta$) and remains almost constant over a wide range
of densities. We note that a direct experimental observation of the
magneto-anisotropy breakdown
 will be difficult as tuning the carrier density by an
external gate voltage will also change the Rashba/Dresselhaus ratio.
\section{Summary}
In conclusion, we have investigated the spin dephasing in a
high-mobility 2DES as a function of the initial spin polarization,
showing that due to electron-electron interaction, the spin
dephasing is reduced as the initial spin polarization is increased.
Additionally, we studied the spin dephasing in external magnetic
fields applied perpendicular and parallel to the 2DES. Here, we
observed a strong in-plane anisotropy of the spin dephasing due to
interference between the Dresselhaus and Rashba spin-orbit fields.
This anisotropy is strongest for equal Dresselhaus and Rashba
fields, yet highly density-dependent, as the \emph{kubic}
Dresselhaus term changes the symmetry of the combined effective
field.
\section{Acknowledgements}
The authors would like to thank J. Zhou, J.L. Cheng, J.H. Jiang and
M.W. Wu for fruitful discussion. Financial support by the DFG via
SPP 1285 and SFB 689 is gratefully acknowledged.

%
%

\end{document}